# Ground-State Probabilistic Logic with the Simplest Binary Energy Landscape for Probabilistic Computing

Yihan He[1], Sheng Luo[1], Chao Fang[1], Gengchiau Liang[1,2]*

[1]Department of Electrical and Computer Engineering, National University of Singapore, Singapore 117576

[2]Industry Academia Innovation School, National Yang-Ming Chiao Tung University, Hsinchu City 300093, Taiwan

Corresponding Author: gcliang@nycu.edu.tw

*Abstract*—*We investigate the ground-state probabilistic logic based on a binary energy landscape (GSPL-BEL) model, implementing the many-body interactions within Ising model cells. The GSPL-BEL model offers a simplified binary energy landscape, enabling the conversion of traditional CMOS-based logic into a probabilistic graphical representation based on desired truth tables. Stochastic Ising cells, coupled with generic probabilistic devices exhibiting sigmoidal electrical responses, serve as the building blocks of the GSPL-BEL. Multi-body interactions are realized through cascaded CMOS-based XNOR gates and a passive resistor network. Through circuit simulations of three-node, four-node, and five-node systems, the functionality of the GSPL-BEL model is verified in forward, reverse, and partial-reverse operating modes, and applied to various arithmetic tasks. The many-body effect provides additional degrees of freedom in describing the system's energy function, resulting in distinct energy levels for valid and invalid states. This observation is supported by the binarized probability distribution observed in the steady state of the probabilistic circuits. Furthermore, compared to conventional combinatorial logic circuits, the GSPL-BEL-based circuit design requires a minimal number of probabilistic devices, as demonstrated in the invertible multiplier/integer* *factorizer circuit. These findings highlight the potential of the GSPL-BEL model for future high-performance logic circuit designs leveraging probabilistic devices.*

I. INTRODUCTION

In recent years, the field of probabilistic computing has seen a surge in research focused on emerging nanodevices, such as low-energy barrier magnetic tunnel junctions [1]–[7]. These devices possess inherent stochasticity and rapid fluctuation capabilities [8], [9], rendering them well-suited for extensive network configurations within innovative computational paradigms. A prominent application in this arena involves the quantum-inspired Ising model/Boltzmann machines-based solver [2], [10]–[13], holding promise for addressing intricate computational challenges, such as integer factorization [14] and Boolean satisfiability [15].

The essence of invertible logic lies in the optimal solution mapped to the energy ground state. Similar to the thermodynamic principles, the probabilistic system would evolve to the ground state as the circuit operates, and hence, it also could be referred to as ground-state probabilistic logic. In physics, the energy landscape delineates the potential energy levels within a system, with the state possessing the lowest energy referred to as the ground state. Likewise, in the ground-state logic, the Boolean function of the target logic can be mapped to the energy landscape through computational models, notably including the Ising model [16] and the Boltzmann machine [17]. Each state within the state space is associated with a distinct energy level, and the objective is to identify the ground state, which signifies the optimal or desired solution. This ground state corresponds to a state configuration that aligns with the truth table. To avoid the system trapped in energy local minima, the perturbations induced by the stochastic devices empower the system to escape these local energy minima, ultimately driving the system toward thermal equilibrium. Consequently, the probabilities associated with all potential solutions will converge to follow a Boltzmann distribution in accordance with their respective energy levels.

Up to now, the ground-state computing model based on many-body interacting spin systems [18] has demonstrated the capability to perform effective error-free computation effectively. This achievement provides a theoretical foundation to implement the ground-state probabilistic logic, but it is without practical implementation designs yet. Currently, the feasibility of hardware implementation for Markov random field-based probabilistic logic circuits including inverters, NAND gates, Half adders, Full adders, etc., have been verified in several works [19]–[23], wherein CMOS-based modules with noise have been employed. Nevertheless, in these works, there is no attention paid to the information of probability distributions resulting from the perspective of the system's energy landscape, as well as the hardware implementation of potential multi-interactions. An algebraic method has been developed to simplify the multi-body interactions to two-body interactions using the same number of spins [24]. While this approach involves the study of the energy landscape and can accurately map the correct solution to the ground state, the resulting energy landscape of the ground-state spin logic becomes more complicated due to its failure to fully capture the wrong solutions. For example, for the ground-state spin AND gate, the number of different energy levels increases from 2 to 3. A similar problem is also evident in recent probabilistic bit (p-bit)-based ground-state probabilistic logics that utilizes mainstream two-body interactions [2], [10], [12], [25], [26]. As the handling of wrong solutions within the solution space remains incomplete, these unwanted states are assigned to multiple discrete energy levels rather than a singular one. Consequently, additional states, identified as "partially wrong" or "significantly wrong", appear in the energy landscape beside the categories of "correct" and "wrong" solutions. This situation worsens when the above basic logic gates are logically synthesized to create combinatorial ground-state probabilistic logic circuits, such as invertible multipliers. Energy levels of building blocks will be stacked in the cumbersome logic synthesis process, leading to a much more complicated landscape. To enhance the performance of these combinatorial circuits with intricate landscapes, resources-consuming annealing techniques such as simulated annealing [7], [27] and parallel annealing [5] are inevitable to employ to facilitate the system's progression toward the ground state. An effort has been undertaken to simplify energy landscapes in CMOS-based ground-state probabilistic logic circuits [28]. However, it focuses only on investigating the effects of three-body interactions to simplify the energy landscape of small-scale logic components, such as the 3-node gates and adders. In addition, this study as well as other previous work [29] which investigates logically invertible probabilistic logic involving higher-order interactions both rely on linear programming [25], [30] to determine the configuration parameters of the logic gates, which incurs additional algorithmic costs and proves impractical as circuit size scales up. Therefore, there is a compelling need for an analytical derivation of ground-state design involving many-body interactions, which could be universally applied to the probabilistic logic systems, irrespective of their scale.

In this paper, we present a computational model for ground-state probabilistic logic based on the Ising model with an optimized energy landscape. This model, known as ground-state probabilistic logic with binary energy landscape (GSPL-BEL), extends the conventional pairwise interactions to multi-body dimensions, significantly simplifying the complexity of the energy landscape. The highest multi-body interaction dimension available in a GSPL-BEL is consistent with the number of nodes in the system. This work, furthermore, demonstrates how the GSPL-BEL model can transform any arbitrary CMOS-based deterministic logic into probabilistic graphical models with the simplest energy binary landscape, facilitating error-tolerant logic operations. To validate the functionality of the proposed GSPL-BEL model, we conducted circuit simulations employing probabilistic devices to emulate the Ising cells. Additionally, we exhibit how many-body interactions can be implemented using conventional XNOR gates, rendering a compatible circuit solution with existing CMOS technology nodes. This advancement opens new possibilities for harnessing the benefits of multi-body interactions in logic operations, further expanding the potential of the GSPL-BEL model in diverse probabilistic computing applications.

## II. GROUND-STATE PROBABILISTIC LOGIC WITH BINARY ENERGY LANDSCAPE

### A. *Computational model based on many-body interactions*

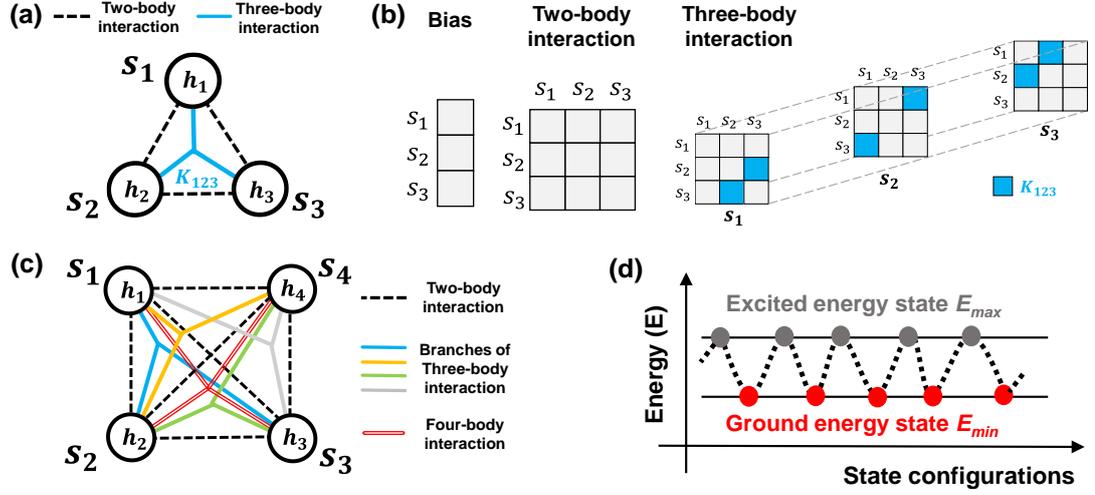

Fig. 1 (a) The general graphical models and (b) the mathematical representations of the three-body-interacting system, which is characterized by the bias term, two-body interaction term, and three-body interaction term. (c) The graphical model of the four-body-interacting system incorporating interaction dimensions up to four-body. (d) Schematic diagram of the logic's binarized energy landscape based on the GSPL-BEL model.

The fundamental physical mechanism behind the GSPL-BEL model is grounded in the Boltzmann Law, wherein the configuration of the advanced Ising model/ Boltzmann machine with multi-body interactions dictates the characteristics and behavior of the system. Fig. 1(a) provides an illustrative graphical model of a three-body-interacting system that fully utilizes higher dimensions of interactions, where each node's state, denoted as *s*, can only take binary values 0 and 1. In contrast to the conventional structure of the Ising model, which features pairwise interactions, the GSPL-BEL model shown in the figure incorporates more intricate interactions. For instance, in this 3-node case, a single branch of three-body interaction is introduced and can be represented in a tensor with $(s_1, s_2, s_3) = (s_1, s_3, s_2) = (s_2, s_1, s_3) = (s_2, s_3, s_1) = (s_3, s_1, s_2) = (s_3, s_2, s_1) = K_{123}$ shown in Fig. 1(b). In the case of the 4-node-interacting system depicted in Fig. 1(c), the model incorporates four branches of three-body interactions (shown in blue, yellow, green, and grey color) and one branch of four-body interactions (shown in hollow red line), therefore, the scope of interactions within the system is further expanded and more design space is provided to achieve more powerful arithmetic operations.

To generalize the GSPL-BEL model, an *M*-dimensional vector, denoted as $\boldsymbol{x}$, with components $x_1, x_2, \ldots x_m$, and an *N*-dimensional vector, denoted as $\boldsymbol{y}$, with components $y_1, y_2, \ldots, y_n$ are defined. A logic function *f* that maps $\boldsymbol{x}$ to $\boldsymbol{y}$ can be represented as $(y_1, y_2, \ldots, y_n) = f(x_1, x_2, \ldots x_m)$, where elements in $\boldsymbol{x}$ and $\boldsymbol{y}$ are binary variables, taking values from the set {0, 1}. The key concept behind designing GSPLs-BEL is that valid states, which conform to the target logic function within the state space, i.e., $S_{y=f(x)} = \{(x_1, x_2, \ldots x_m, y_1, y_2, \ldots, y_n) \mid f(x_1, x_2, \ldots x_m) = (y_1, y_2, \ldots, y_n)\}$ should configure the system to the ground state $E_{min}$. In contrast, other invalid states $S_{y \neq f(x)} = \{(x_1, x_2, \ldots x_m, y_1, y_2, \ldots, y_n) \mid f(x_1, x_2, \ldots x_m) \neq (y_1, y_2, \ldots, y_n)\}$ contribute a penalty energy $E_{max}$ to the system, as follows:

$$E = \begin{cases} E_{min}, & S_{y=f(x)} \\ E_{max}, & S_{y \neq f(x)} \end{cases} \quad (1)$$

It is important to note that $E_{min} < E_{max}$. By fully harnessing the capabilities of many-body interactions, this model can achieve the utmost simplicity in the energy landscape for the target logic, resulting in the simplest binary energy landscape with two distinct energy levels, as illustrated in Fig. 1(d). When the system reaches thermal equilibrium at a finite temperature, the steady-state probability for a specific state configuration can be described by the Boltzmann Law:

$$P(\{s\}) = \frac{\exp\left(-\frac{E(\{s\})}{T}\right)}{\sum_{i,j} \exp\left(-\frac{E(\{s\})}{T}\right)} \quad (2)$$

where $T$ is a pseudo-temperature parameter that reflects the degree of stochasticity of the system in the context of ground-state probabilistic logic. In the subsequent section, we will introduce how each Ising cell is implemented with a probabilistic device.

Following this, by substituting the binarized energy levels, i.e., $E_{min}$ and $E_{max}$ into Eq. (2), we can compute the theoretical solutions for the statistical probabilities of valid and invalid states as follows:

$$P(S_{y=f(x)}) = \frac{\exp\left(-\frac{E_{min}}{T}\right)}{N_{S_{y=f(x)}} \cdot \exp\left(-\frac{E_{min}}{T}\right) + N_{S_{y \neq f(x)}} \cdot \exp\left(-\frac{E_{max}}{T}\right)} \quad (3a)$$

$$P(S_{y \neq f(x)}) = \frac{\exp\left(-\frac{E_{max}}{T}\right)}{N_{S_{y=f(x)}} \cdot \exp\left(-\frac{E_{min}}{T}\right) + N_{S_{y \neq f(x)}} \cdot \exp\left(-\frac{E_{max}}{T}\right)} \quad (3b)$$

where $N_{Sy = f(x)}$ and $N_{Sy \neq f(x)}$ are the number of valid and invalid states in the state space, respectively.

A simple three-cell logic system is used to illustrate the procedure. The configuration of this three-body interacting model is characterized by a set of connectivity parameters, denoted as $\{h_A, h_B, h_C, J_{AB}, J_{AC}, J_{BC}, K_{ABC}\}$. We will use an AND gate with input nodes $s_1$ and $s_2$, and an output node $s_3$, to demonstrate the process of deriving the configuration parameters that can encode the simplest energy landscape. In the first step, summing all clique energies $E_{s1s2s3}$ in the state space:

$$\begin{aligned} E_{f(AND)} &= \sum E_{S_{y=f(AND)}} + \sum E_{S_{y \neq [f(AND)]}} \\ &= \{E_{min(000)} + E_{min(010)} + E_{min(100)} + E_{min(111)} \\ &\quad + E_{max(001)} + E_{max(011)} + E_{max(101)} + E_{max(110)}\} \end{aligned} \quad (4)$$

The total energy of the system needs to be binarized to $E_{min}$ and $E_{max}$, after applying constraints. Specifically, when the state corresponds to one of the four valid configurations $(s_1\ s_2\ s_3) = (0\ 0\ 0)$, $(0\ 1\ 0)$, $(1\ 0\ 0)$, and $(1, 1, 1)$, the energy of that state configuration is set to the ground state $E_{min}$. On the other hand, for the other 4 undesirable states, the system is configured to have an energy level of $E_{max}$. To express the energy in terms of variables that meet these constraints, we employ Boolean ring conversion and represent binary states 0 and 1 as $1-s$ and $s$, respectively:

$$\begin{aligned}E_{f(x)\leftrightarrow AAND} &= E_{min}[(1-s_1)(1-s_2)(1-s_3) + (1-s_1)(s_2)(1-s_3) \\ &\quad + (s_1)(1-s_2)(1-s_3) + (s_1)(s_2)(s_3)] \\ &\quad + E_{max}[(1-s_1)(1-s_2)(s_3) + (1-s_1)(s_2)(s_3) + (s_1)(1-s_2)(s_3) \\ &\quad + (s_1)(s_2)(1-s_3)] \\ &= (E_{max} - E_{min})s_3 + (E_{max} - E_{min})s_1s_2 - 2(E_{max} - E_{min})s_1s_2s_3 + E_{min} \quad (5)\end{aligned}$$

Because the Ising model uses a bipolar representation format for variables, the energy function is then converted from the binary format using $s = (m+1)/2$:

$$E_{AND} = \frac{1}{4}(E_{max} - E_{min})\left[m_3 - m_1m_3 - m_2m_3 - m_1m_2m_3 + 2\frac{E_{max} + E_{min}}{E_{max} - E_{min}}\right] \quad (6)$$

where $m$ represents the bipolar value $-1$ and $+1$. Meanwhile, the energy of a many-body interacting Ising system is defined as follows:

$$E(\{s\}) = -\left(\sum_i h_i m_i + \sum_{i<j} J_{ij} m_i m_j + \sum_{i<j<k} K_{ijk} m_i m_j m_k \right. \\ \left. + \sum_{i<j<k<l} L_{ijkl} m_i m_j m_k m_l + \cdots \right) \quad (7)$$

The energy of this 3-node AND gate is:

$$E_{AND} = -(h_1 m_1 + h_2 m_2 + h_3 m_3) - (J_{12} m_1 m_2 + J_{13} m_1 m_3 + J_{23} m_2 m_3)$$

$$-K_{123}m_1m_2m_3 \tag{8}$$

The configuration parameters of the system can be determined by computing the ratio of coefficients for the bias $\{h\}$, pairwise $\{J\}$, and three-body term $K_{123}$. The constant term in the calculation exactly corresponds to the ground state of the system. This approach is applicable to expand to any target logic, as long as its truth table is known. Fig.2 presents the pseudo-code for translating conventional CMOS logic into corresponding graphical models using this model, offering a comprehensive framework for designing various logic functions.

```
Parameters:
    Truth_Table [T]           // Truth table for the logic problem to be solved
    Parameters [s]            // Binary representation of states
    Parameter n               // Width of Truth Table
Begin
    For i ∈ {1 ··· n}
        S[i,1] ← 1            // Recording of constant terms
        S[i,2] ← 0.5          // Recording of primary term coefficients
    EndFor

    For i ∈ {1 ··· 2^n}       // Repeat length of Truth table times
        For k ∈ {1 ··· n}     // Repeat width of Truth table times
        If Truth_table(i,j)==0
        Then
            Ans(i,j,1) ← 1-s(j,1)
            Ans(i,j,2) ← 1-s(j,2)    // 0 → 1-s
        Else
            Ans(i,j,1) ← s(j,1);
            Ans(i,j,2) ← s(j,2);     // 1 → s
        end
    EndFor
    Answer, Position ← multiply(Ans) // multiply is a function to calculate the product result of Ans,
                                     which can finally get the values and positions of coefficients
    EndFor
End
```

Fig. 2 Pseudo-code for converting arbitrary traditional CMOS-based deterministic logic into GSPL-BEL represented by graphic models. The conversion is based on the truth table of the target logic.

### B. *Examples of GSPL-BEL using graphic representations*

We have presented a general framework for determining the connectivity of GSPLs-BEL by designing energy functions. To demonstrate the operation of the model, we will investigate small-scale GSPLs-BEL with the number of nodes ranging from 3 to 5 and provide their graphical representations. A comprehensive GSPL library has been established, which can be directly utilized for further studies.

The majority gate serves as a core component in various applications, including image processing [31] and brain-inspired computing [32]. We explore a potential implementation of a majority gate based on the GSPL-BEL model. After normalizing the connectivity strength by setting $E_{min}$ and $E_{max}$ in Eq. (1) to 0 and 1, respectively, the energy function of a 3-input majority gate composed of nodes $A$, $B$, $C$, and $O$, can be described as follows:

$$E_{Majority} = -m_A m_O - m_B m_O - m_C m_O - m_A m_B m_C m_O - 2. \qquad (9)$$

The absolute values of the coefficients in the first three terms and the fourth term define the interaction strength of pairwise and four-body connectivity, respectively. The overall configurations can be captured by the graphical representation shown in Fig. 3(a). This model is composed of four nodes and the sign of interaction strength represents the direction of interaction received from neighboring cells, for instance, cell $A$ receives a negative four-body-interaction signal from node $B$, $C$, and $O$.

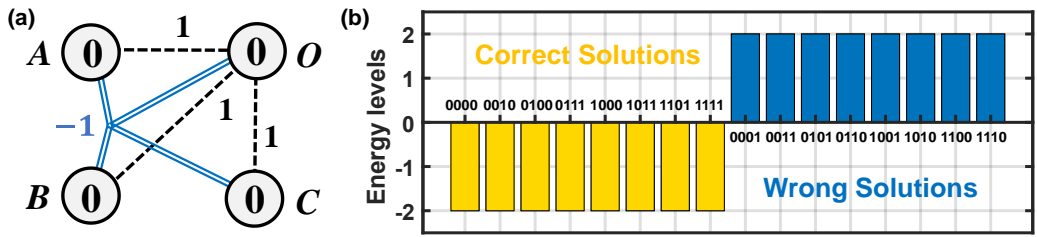

Fig. 3 (a) The graphic representation of the 3-input and 1-output majority gate. The sign of integers marked on the connection lines indicates the polarity of interactions received from other nodes, which can be negative or positive effects. (b) Binary energy landscape with two energy levels.

The clique energy, $E_{ABCO}$, for each state configuration in the state space of this 4-node majority gate, can be computed using Eq. (7). As illustrated in Fig. 3(b), there are two distinct energy levels, in which all valid and invalid states are degenerate and valid states configure the system to $E_{min} = -2$. It matches with the constant term in Eq. (9). In contrast, the invalid states are all mapped to the maximum energy level, $E_{max}$, which is at +2. As expected, the excited states corresponding to a higher energy will have a lower probability of being occupied at thermal equilibrium. Therefore, we can expect to obtain a binarized probability landscape that depends on the state configuration in this majority gate.

To further illustrate the capabilities of the GSPL-BEL model, additional examples of logic families that are suited for fault-tolerant arithmetic operations ranging from 3-node to 5-node are provided in Fig. 4. By customizing the configuration parameters based on specific truth tables, firstly, various logic functions are designed within three-body-interacting systems, including AND, OR, and XOR, and XNOR. If an additional node is introduced, more complex operations, such as the majority function which we discussed earlier, 3-input and 1-output arithmetic AND operation, half addition operation can be achieved by incorporating the four-body interactions into the 4-node network. Furthermore, the half addition operation can be upgraded to a full addition operation by adding one more node to serve as the carry-in node $Ci$. This 5-body-interacting Full adder can propagate any carry generated from lower-order bits to higher-order bits, making it suitable for multi-digit binary addition. Under our proposed GSPL-BEL model, these small-scale many-body-based logic components all exhibit a binarized energy landscape and could serve as building blocks of combinatorial probabilistic logic circuits. For example, by combing the many-body-based AND gates, Half adders, and Full adders, integer factorizers can be constructed to solve the integer factorization problem. Besides, NOT gate, three-body-based AND gate, and OR gate can be logically synthesized to the solver of Boolean satisfiability [33]. The simplification effects of the many-body-based design on logically synthesized factorizers have been investigated in our previous research [29]. However, it is imperative to note that the development of probabilistic models for larger-size logic circuits in this work does not involve a logic synthesis process, as the combination of basic gates would introduce additional and unnecessary energy levels, thereby significantly complicating the energy landscape. Besides, the solution space of the factorization problem will become larger, due to the involvement of auxiliary nodes. Our primary objective is to present a design solution that achieves the simplest energy landscape for ground-state probabilistic logics of any size with minimal overhead of nodes and validate the functionality of GSPL-BEL using probabilistic devices.

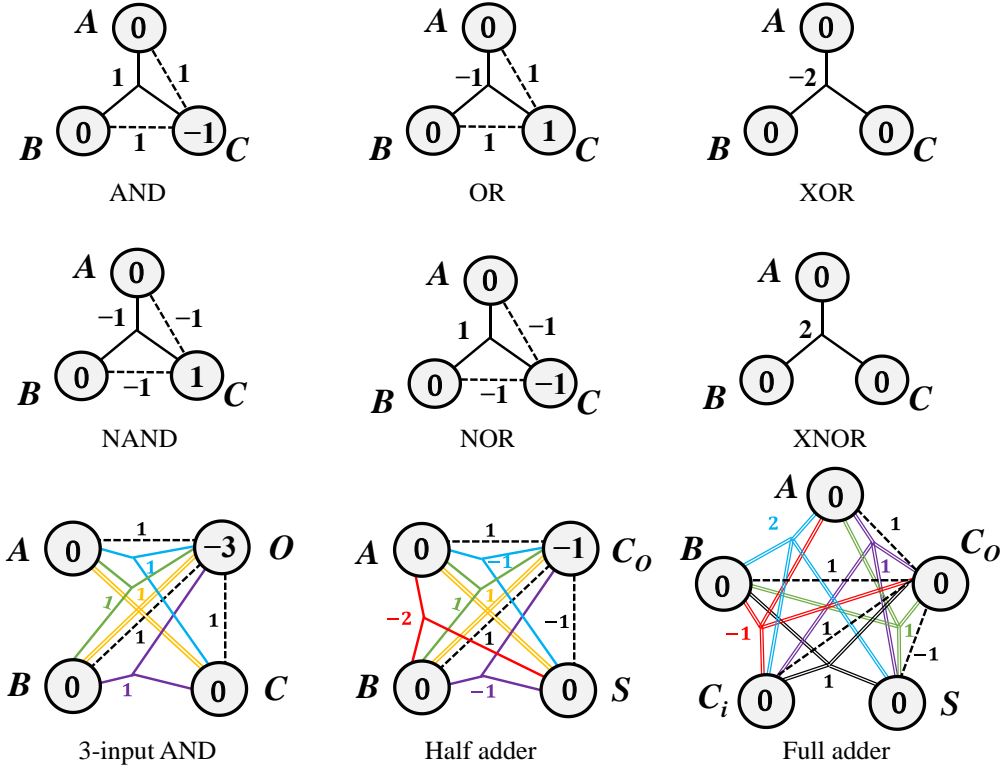

Fig. 4 Configuration library of many-body-interacting fundamental GSPL-BEL gates using graphic representation.

## III. HARDWARE IMPLEMENTATIONS

### A. *Generic probabilistic device*

In this section, the processes of translating the graphical representations of GSPLs-BEL into electronic elements will be described in detail. A generic three-terminal probabilistic device with an analog input terminal and a digital output terminal is shown in Fig. 5(a), which implements the cell of GSPL-BEL. As depicted, the output signal from this device is a binary voltage, accepting only two values, 1 and 0, representing the high voltage level $V_{DD}$ and the low voltage level 0, respectively. This device is capable of constantly generating fluctuating bitstreams of 1s and 0s. By adjusting the direction and strength of the input signal, the probability of producing 1s can be tunable.

This stochastic behaviour follows a sigmoidal relation and can be described as follows [2]:

$$s_i(t) = sgn\{rand(-1, 1)\} + \tanh[I_{in}(t)]\} \quad (10)$$

where $I_{in}$ is a current signal and rand $(-1, +1)$ is a uniformly distributed random number between $-1$ and $+1$.

The probability of an output of 1 in response to the input signal is obtained statistically by averaging the output values over time. So far, a variety of probabilistic devices have demonstrated their functionalities in the field of probabilistic computing. These devices include noise-based standard CMOS logic gates [21]–[23], thermal noise-driven low-energy-barrier magnetic tunnel junctions [3], [34], programmed microcontrollers [35], CMOS-based [12] and FPGA-based [26] probabilistic circuits using pseudo-random bitstreams, as well as other emerging probabilistic devices [36], [37]. Networks and systems built upon these devices have proven effective in solving a range of hard computational problems, including integer factorization [2], [10], combinatorial optimization problems [38], Bayesian inference [6], and machine learning [39]. However, in order to minimize the time required to reach solutions, the fluctuation time between 0s and 1s of probabilistic devices should be as short as possible, while still adhering to the limits imposed by the circuit design [40]. This is also a key design metric for the GSPL-BEL. In this work, we start by characterizing the probabilistic characteristics of a given probabilistic device through a fitted sigmoidal curve shown in red color or a lookup table that reflects the discrete data points in black, as shown in Fig. 5(b). Subsequently, we develop a Verilog-A behavioral model in Cadence Virtuoso after modeling and packaging the device to a modular cell. Finally, the cells are assembled into probabilistic logic with various computational functions, based on a pre-designed network structure that incorporates many-body effects. Utilizing probability statistics grounded in Boltzmann's law, we evaluate and analyze the overall performance of the GSPL-BEL at the circuit level. The implementation of many-body effects based on CMOS electronic elements will be discussed in the next section.

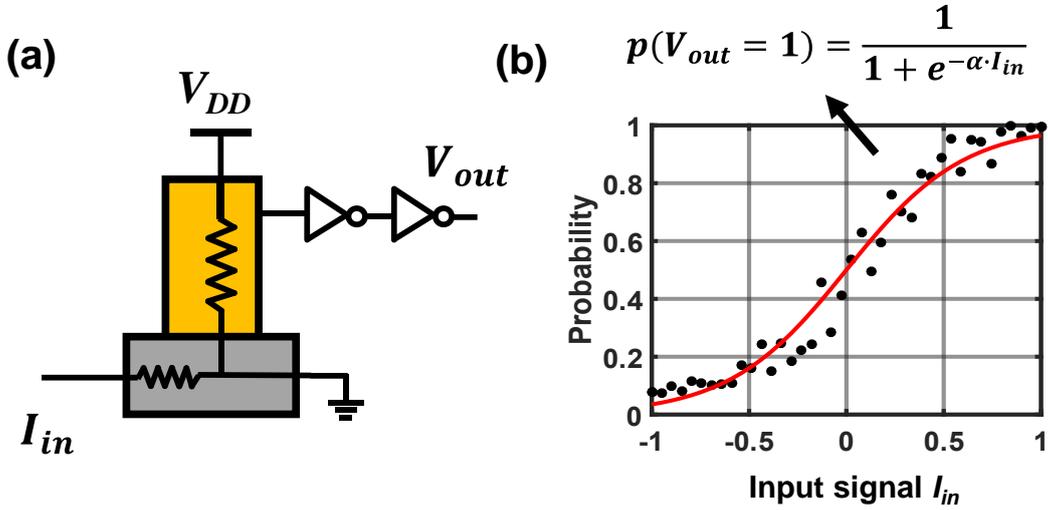

Fig. 5 (a) Generic probabilistic building block for GSPL-BEL with analog input and digital output. (b) Sigmoidal response of the probabilistic device with respect to the input signal, where α is a parameter reflecting the tilt degree of the S-shape curve and it is determined by the stochastic behavior of probabilistic devices.

B. *Many-body interactions*

On the basis of the use of probabilistic devices to implement Ising cells, another critical step in designing many-body interacting GSPLs-BEL is to determine suitable hardware implementations for multi-order interactions among cells. Previous work of a probabilistic computer [1] first leverages the many-body interactions among p-bits, but the interactions are realized through customized programs and subsequently executed by peripheral microcontrollers, resulting in substantial hardware overhead.

To streamline the design of network connections, we choose a probabilistic device with an analog current input and digital voltage output to function as the Ising cell. This kind of device can make the hardware implementation of interactions within two dimensions simpler and more straightforward, as the conversion from the voltage output of the cell to the current input of neighboring interacting cells can be easily implemented through a passive resistor network [2]. The feedback current serves as the carrier of the cell-to-cell interaction. During operation, cells are updated sequentially, and the strength of incentive for cell $m_i$ depends on the cumulative interaction values of its neighboring cells:

$$I_i(t) = h_i + \sum_j J_{ij} s_j(t) + \sum_{j,k} C_{ijk} s_j(t) s_k(t) + \sum_{j,k,l} D_{ijkl} s_j(t) s_k(t) s_l(t) + \cdots \quad (11)$$

Only considering the third term representing the three-body interaction and an $N$-body interaction, the contribution of one of their respective branches to the total accumulated current is as follows:

$$I_{i,3-body} = C_{123} s_1 s_2 \quad (12a)$$

$$I_{i,N-body} = Z_{N-body} \cdot \prod_{1}^{N-1} s_i \quad (12b)$$

where $Z_{N\text{-}body}$ represents the coefficient for the $N$-body interaction. In the top left of Fig. 6(a), the theoretical values of nodes $m_1$, $m_2$, and $m_1 \cdot m_2$ after the three-body interaction are represented in the bipolar format. To implement the circuit in practice, these values must be converted to binary format using the function $f(s_1,s_2)$, as the output information of cells is encoded using digital 0 and 1 in the circuit. Interestingly, this function perfectly matches the operation of a conventional XNOR gate in the bottom left. Furthermore, by cascading $N-2$ XNOR gates in series, the hardware implementation of many-body interactions can be scaled up to an $N$-body system. Fig. 6(b) illustrates this implementation, where the output signals of cells $s_1$ and $s_2$ are first processed by the first XNOR gate and then fed to subsequent XNOR gates. The final output signal following the $N$-body interaction is obtained from the output of the last XNOR gate, namely the $(N-2)$th XNOR gate.

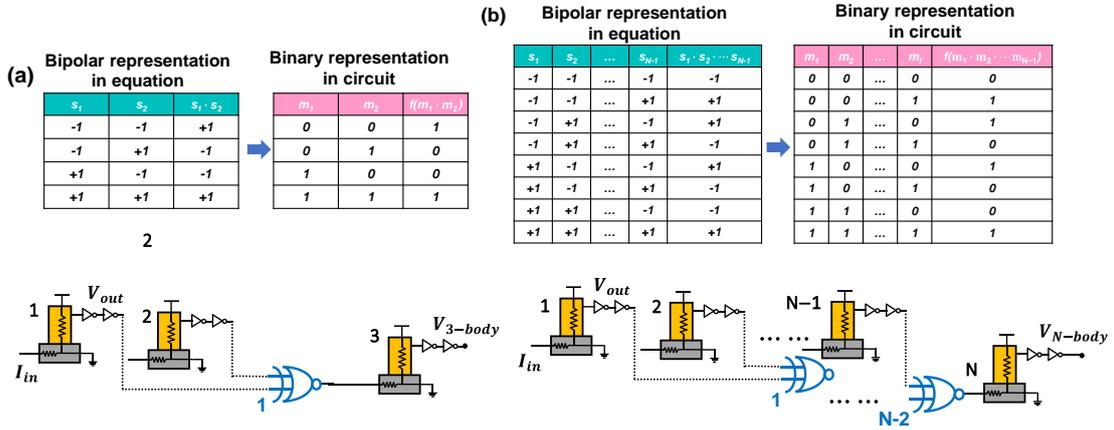

Fig. 6 (a) The differences between mathematic and circuit representations of the three-body effect. One XNOR gate in blue color is used to implement the three-body interaction. (b) The difference between mathematic and circuit representations of the four-body effect. $N$-2 serially connected XNOR gates are used to implement the $N$-body interaction.

## IV. RESULTS AND DISCUSSION

### A. *Example of the GSPL-BEL with Electronic Elements*

Fig. 7 presents a 4-node Majority gate designed based on the proposed GSPL-BEL model, where all graphical representation information has been translated into electronic components. Specifically, we combine a resistor network with XNOR gates to implement the four-body interactions. We make several simplifying assumptions in the circuit implementation: 1) The resistance of the underlying path labeled as $R_{UL}$ is considered negligible. As a result, the voltage of the input terminal for each cell can be fixed to $V_{DD}/2$, which is nearly equal to the bias voltage applied at the third terminal. As a result, when the input current equals to 0 ($V_{in} = V_{DD}/2$), a 50% probability of getting 1 can be obtained. 2) The response time, which includes the total time for cell retention and fluctuation, is assumed to be approximately 1 ns [8], [9]. 3) The transmission time of XNOR gates used to implement many-body interactions is much shorter than the response time of cells and there is no delay across CMOS gates. Circuit simulations are conducted using the HSPICE simulator in Cadence Virtuoso, with detailed parameters summarized in Table I.

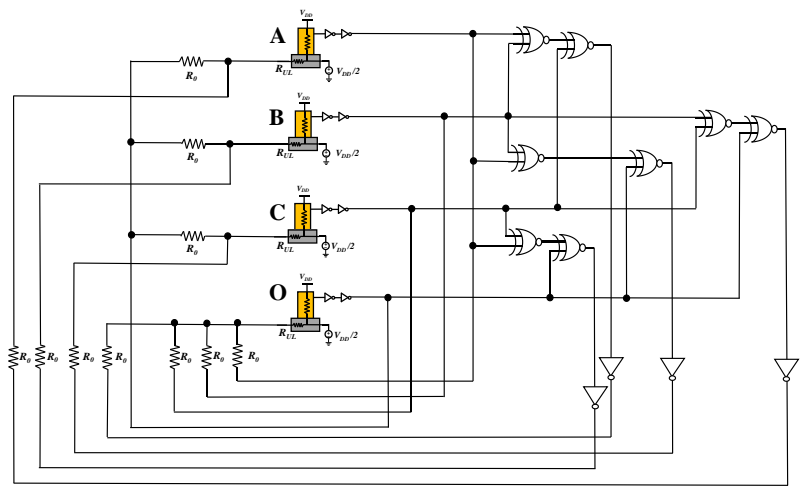

Fig. 7 Schematic of the GSPL-BEL-based majority gate, in which the interactions are implemented in hardware with resistors together with conventional XNOR gates.

TABLE I  PARAMETERS USED IN CIRCUIT SIMULATIONS

| Electronic elements | Parameters | Value |
| --- | --- | --- |
| Probabilistic building block | Tilt degree parameter $\alpha$ | $4 \times 10^6$ |
| | Fluctuation time (ns) | 1 |
| Resistor | $R0$ ($\Omega$) | $1 \times 10^6$ |
| Voltage source | Global voltage (V) | 1 |

In the absence of terminal constraints, the majority gate operates in a free mode where the four constituent cells continuously fluctuate. This dynamic behavior enables the majority gate to explore all $2^4$ possible state configurations within its state space. During operation, the probabilistic circuits will spend more time on states with lower energy, as evidenced by the statistical results shown in Fig. 8(a). The observed phenomenon of probability binarization in the figure indicates that our model encodes the energies of valid states and undesirable states into two distinct levels at $-2$ and $+2$ according to Eq. (7). Furthermore, by clamping different terminals, GSPL-BEL can operate in three modes to handle various tasks. For example, a GSPL-BEL-based adder/ multiplier can separately function as a subtractor/ divider, and a sum factorizer/ integer factorizer by operating in the partial forward mode and reverse mode, respectively. Considering a 5-node full adder, in the forward mode, this probabilistic model functions similarly to conventional CMOS-based logic. When $A$, $B$, and $C_{in}$ are clamped to 0, 1, and 1 respectively, ($S$, $C_{out}$) exhibits a 99.90% probability of producing the correct solution (0, 1). Unlike the deterministic operations of CMOS, the results of GSPL-BELs' operations are achieved through probabilistic statistics. On the other hand, the reverse operation can be activated by clamping the output terminal. For example, if the terminals $S$ and $C_{out}$ are clamped to 0 and 1, respectively, then ($A$, $B$, $C_{in}$) will generate all state combinations that satisfy this addition result with high probability. The probabilities of correct solutions ($A$, $B$, $C_{in}$) = (0, 1, 1), (1, 0, 1), and (1, 1, 0) are approximately 33.28%. Finally, by clamping all the output terminals ($S$, $C_{out}$) to (1, 0) and some of the inputs ($A$, $C_{in}$) to (0, 0), the full adder can operate in the partial forward mode to function as a subtractor. Similarly, this encoding scheme can effectively distinguish favorable and unfavorable states of the full adder operating freely, as demonstrated in Fig. 8(b). The binary energy landscape aids in visualizing and analyzing the probabilities associated with different states.

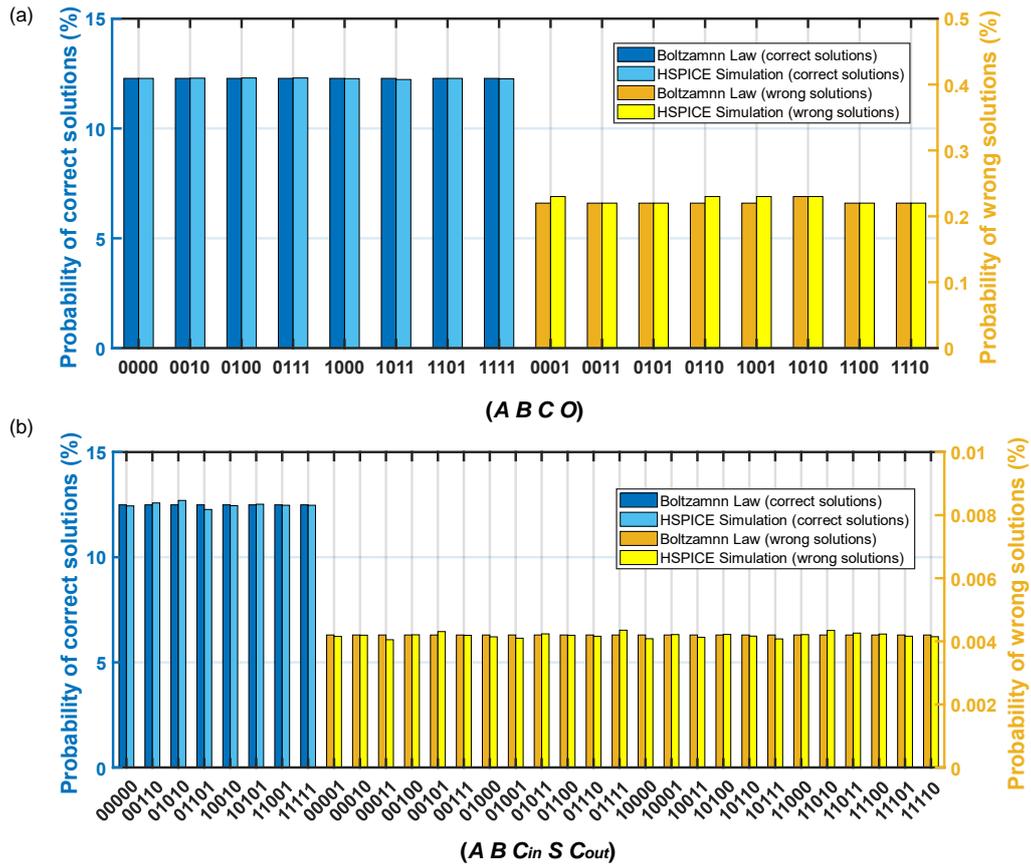

Fig. 8 The statistical probability distributions of the 4-node majority gate and 5-node full adder based on the GSPL-BEL model. Statics of both cases are obtained by averaging $2 \times 10^6$ sampling points in the time domain.

The above two examples demonstrate the fault tolerance of probabilistic logic operations, as the result of an arithmetic operation is not determined by a single node state configuration at a certain moment but depends on the probability distribution of all possible state configurations over a defined time interval. This signifies that even if the system momentarily enters an incorrect solution state, it possesses the capacity to transition out of the current state—a capability unattainable within traditional CMOS logic, which performs operations deterministically.

*B. Non-logic synthesis GSPL-BEL*

To comprehensively evaluate the functionality of the GSPL-BEL model in a broader range of logic circuits, we explore multipliers of various sizes. As mentioned earlier, GSPL-BEL-based multipliers have the ability to operate in the reverse mode, functioning as an invertible multiplier or an integer factorizer. This unique feature has promising applications in encryption and machine learning.

Large-scale invertible multipliers can be constructed using logic synthesis from two-body-based basic gates [12], GSPL-BEL model-based basic gates [29] (AND, Half adder and Full adder shown in Fig. 4) or through direct implementation using the GSPL-BEL model itself. In Fig. 9(a), a general $N$-bit adder-based multiplier created using the logic synthesis method is shown, illustrating the key components and architecture of the multiplier, which includes $N^2$ AND gates, $N$ Half adders, and $N(N-2)$ Full adders, with a total consumption of $3N^2$ nodes. The connection of the basic gates is facilitated by merging the common nodes of basic gates, as exemplified in Fig. 9(b), in which nodes in white color are the common nodes of a 2-bit×2-bit invertible multiplier. These auxiliary nodes serve as bridges to connect different gates. Green and orange nodes are traditionally defined as input and output nodes respectively. In contrast, the GSPL-BEL-based model offers an alternative to bypass the conventional framework of constructing combinational logic circuits through the logic synthesis process, which can be cumbersome and time-consuming. Specifically, this approach allows for the direct design of the energy function based on the multiplier's truth table, followed by the determination of the network configuration parameters. As a result, the GSPL-BEL model provides the most compact design for the multiplier with minimal overhead, with the number of nodes solely determined by the sum of input and output terminals. In contrast to the polynomial growth observed in the conventional logic synthesis method, the GSPL-BEL-based model exhibits linear node expansion relative to the multiplier's size, as illustrated in Fig. 9(c). A paramount advantage stemming from the reduction in required nodes is the significant contraction of the solution space. For instance, in the case of a 2-bit×2-bit invertible multiplier operating in a free mode, this reduction can be achieved from $2^{12}$ to $2^{8}$.

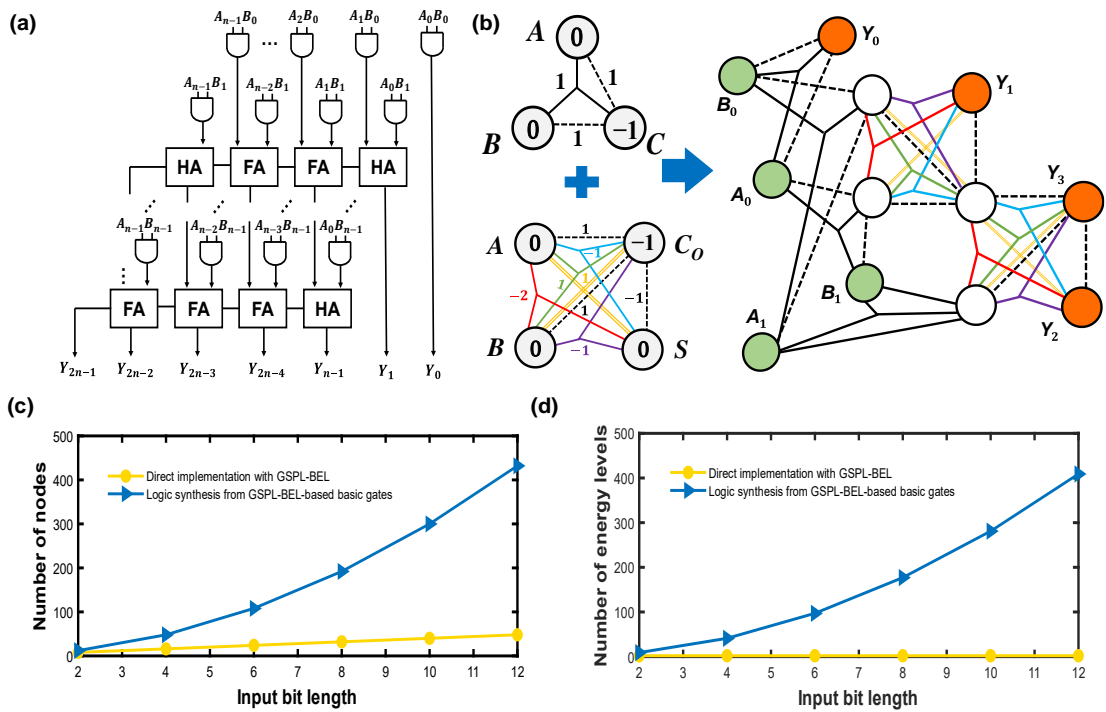

Fig. 9 (a) Schematic of the adder-based invertible multiplier/integer factorizer. (b) A 2-bit×2-bit invertible multiplier logically synthesized from 4 AND gates and 2 Half adders. (c) Comparison of number of nodes. (d) Comparison of number of different energy levels.

Moreover, in the case of direct implementation of the multiplier using the GSPL-BEL model, even though the size of the multiplier grows linearly from 2-bit, the number of energy levels can be maintained at 2, as shown in Fig. 9(d). This is because the dimensionality of the available many-body effects also increases correspondingly with more nodes resulting from the growth in the size of the multiplier. Consequently, for multipliers of any size, the energy can be effectively mapped to only two discrete energy levels, thereby yielding the simplest energy landscape. When comparing multipliers with the same number of input-bit lengths, it can be observed that the multiplier synthesized from GSPL-BEL-based basic gates exhibits a more intricate energy landscape, accompanied by a higher number of nodes, compared to the non-logic synthesis approach. Finally, we compare the factorization accuracy of a 2-bit × 2-bit multiplier with the output clamped to 6 in the reverse mode. Among the three implementations, the non-logic synthesis method provides the highest accuracy for the solutions $(A, B) = (2, 3)$ and $(3, 2) = \sim 50\%$ calculated by Eq. (3), whereas the accuracy for the logic synthesis approach based on two-body-based basic gates is 37.80%. The reason for this improved accuracy of factorization is that the non-synthetic GSPL-BEL model maps the correct and incorrect solutions to two energy levels with a large energy difference, namely −120 and +8.

## V. CONCLUSION

In this study, we introduce a GSPL-BEL model that incorporates many-body interactions, grounded in principles of ground-state computation and energy minimization. This novel model extends the interaction dimensionality within the Ising model, transitioning from two-body to multi-body interactions. This expansion provides greater freedom in shaping the energy function of the system. We illustrate a robust implementation of these many-body interactions through a simple cascade arrangement of conventional XNOR gates. Theoretical calculations based on Boltzmann's law and statistical circuit simulations validate the GSPL-BEL model's efficiency across various functions and sizes. Notably, it offers a compact design solution for probabilistic logic by minimizing node consumption, bypassing conventional logic synthesis. Furthermore, the model achieves binary simplification of energy landscapes for arbitrary logic by harnessing the potential of many-body interactions. Looking ahead, our design showcases promise as a multifunctional, efficient, and fault-tolerant computational model for probabilistic applications.

ACKNOWLEDGEMENT

This work was supported by MOE-2019-T2-2-215 and FRC-A-8000194-01-00. GC. Liang would like to thank financial support from the Co-creation Platform of the Industry-Academia Innovation School, NYCU, under the framework of the National Key Fields Industry-University Cooperation and Skilled Personnel Training Act, from the Ministry of Education (MOE) and industry partners in Taiwan.